\documentclass[a4paper,aps,groupedaddress,showpacs,twocolumn]{revtex4}
\usepackage[dvips]{graphicx}
\usepackage{amssymb}
\usepackage{natbib}
\usepackage{color}
\definecolor{blue}{rgb}{0,0,0}
\newcommand{\ped}[1]{\ensuremath{_{\rm #1}}}

\begin{document}

\title{Point contact spectroscopy in Fe-based superconductors: recent advancements and future challenges}
\author{R.S. Gonnelli, D. Daghero, M. Tortello}
\affiliation{Dipartimento di Scienza Applicata e Tecnologia (DISAT), Politecnico di Torino, 10129 Torino, Italy}
\begin{abstract}
Iron-based superconductors (FeSC) present an unprecedented variety of features both in the superconducting and in the normal state. Different families differ in the value of the critical temperature, in the shape of the Fermi surface, in the existence or absence of quasi-nesting conditions, in the range of doping in which the antiferromagnetic (AFM) and the superconducting phase coexist and in the structure of the order parameter in the reciprocal space, and so on. In this paper the most important results of point-contact spectroscopy (PCS) in Fe-based superconductors are reviewed, and the most recent advances are described with the aim to discuss the future perspectives and challenges of this spectroscopic technique in the characterization of the {\color{blue}{superconducting properties of these complex compounds}}. One of the main challenges, faced so far only by a few researchers in the PCS field, is to fully explore the phase diagram of these materials, as a function of doping or pressure, to understand the interplay between superconductivity and magnetism, the effect of intrinsic or extrinsic inhomogeneities, the role of spin fluctuations (SFs) in the pairing, the symmetry {\color{blue}{and the structure}} of the order parameter(s).
\end{abstract}

%
 \maketitle

\section{Introduction}
\label{sect:intro}
Since its discovery in 1974 \cite{yanson74}, point-contact spectroscopy has been extensively used to study all kinds of scattering of electrons by elementary excitation in metals, like phonons, magnons and so on \cite{naidyuklibro}. In superconductors, thanks to the quantum phenomenon called ``Andreev reflection'', this technique also allows directly determining the amplitude and symmetry of the superconducting order parameter. Point-contact Andreev-reflection spectroscopy (PCARS) has been successfully used in most of conventional and unconventional superconductors, including cuprates, heavy fermions, borocarbides, MgB$_2$ and, since 2008, Fe-based compounds (FeSC) \cite{daghero11}. In the latter case, most of the PCARS experiments carried out up to now have been focused on the determination of the number, the amplitude and the symmetry of the superconducting gap(s) in a relatively small number of representative compounds. However, the capabilities of PCS and PCARS can be exploited much more extensively to study various aspects of the physics of FeSC, that concern their whole phase diagram.
In the following we will present some recent advancements in PCARS that, we believe, can be fruitfully used and further developed to fully exploit this technique in the study of FeSC and in particular to address the open issues concerning: i) the fine structure of the order parameter and its evolution with doping, in connection with the changes in the geometry of the Fermi surface; ii) the pairing mechanism in these compounds, and the determination of the effects of doping on the spectrum of the relevant boson; iii) the  interplay between magnetic order and superconductivity in the region of superposition in the phase diagram of most compounds; iv) the evolution of the static magnetism (i.e. the SDW gap) on doping, and its possible connection with some anomalous features observed in underdoped compounds.

{\color{blue}{This paper is focused on the superconducting properties of FeSC, but PCS has been recently used also to investigate the properties of the normal state. There are very few examples of temperature and/or magnetic field dependence of the point-contact conductance in the normal state of FeSC;  moreover, most of them are poorly or partially discussed in the relevant papers, which are mainly devoted to other topics (order-parameter symmetry, electron-boson features, etc.). In a recent paper, H. Z. Arham and coworkers \cite{arham12} have presented and discussed a large number of PCS results obtained in single crystals of the parent compounds of the 122 family (Ba-122, Sr-122, Ca-122), of doped Ba-122 (both Co- and K-doped) and of $\mathrm{Fe_{1+y}Te}$.  The conductance curves have been collected in a broad temperature range (from few K up to about 200 K) in order to cover the full range of all the possible phase transitions present in the samples: structural (described by the temperature $T_S$), magnetic (described by $T_N$) and superconducting ($T_c$). The results of this investigation are the subject of a contribution by H. Z. Arham and L. H. Greene in this issue of Current Opinion of Solid State and Materials Science.}}

\section{PCARS in the superconducting state}\label{sect:PCARS}
\subsection{Open questions about the order parameter(s)}\label{subsect:OP}
\begin{figure}
\includegraphics[width=0.9\columnwidth]{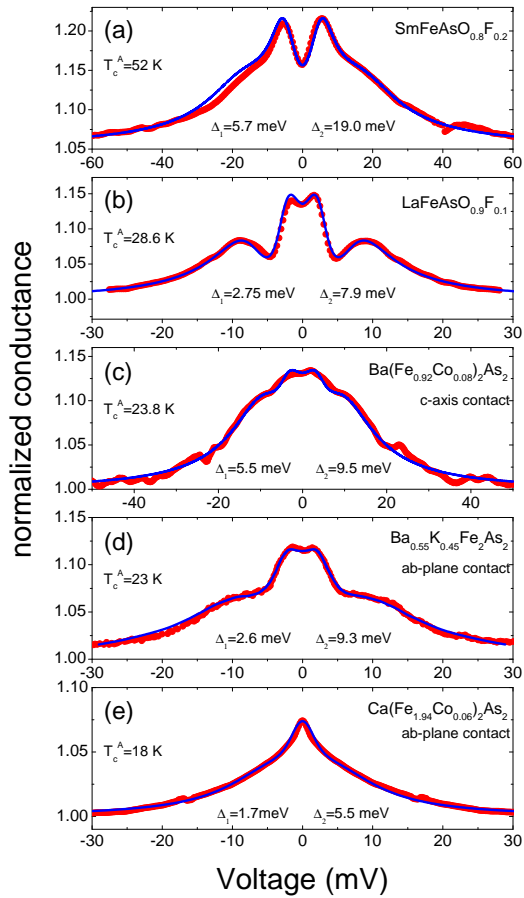}
\caption{Some examples of low-temperature PCARS spectra measured in compounds of the 1111 and of the 122 family of Fe-based superconductors, and more specifically in $\mathrm{SmFeAsO\ped{0.8}F\ped{0.2}}$ polycrystals (panel (a), from \cite{daghero09b}), $\mathrm{LaFeAsO\ped{0.9}F\ped{0.1}}$ polycrystals (panel (b), from \cite{gonnelli09a}), $\mathrm{Ba(Fe\ped{0.92}Co\ped{0.08})_2As_2}$ epitaxial films (with $c$-axis current injection, panel (c)), $\mathrm{(Ba\ped{0.55}K\ped{0.45})Fe_2As_2}$ single crystals (with $ab$-plane current injection, panel (d), from \cite{szabo09}), and $\mathrm{Ca(Fe\ped{0.94}Co\ped{0.06})_2As_2}$ single crystals (with $ab$-plane current injection, panel (e), from \cite{gonnelli12}). Symbols represent the experimental data, solid lines a fit with the two-band, 2D BTK model. In panels (a) - (d), the fit was possible by using two nodeless, isotropic gaps; in panel (e) the fit was obtained by using a fully anisotropic small gap (with zeros, but no sign change) of the form $\Delta_1(\theta)= \Delta_1 \cos^4[2(\theta-\alpha)]$ and a isotropic large gap $\Delta_2$. }\label{fig:symmetry}
\end{figure}

Most of the PCARS experiments in FeSC have been devoted to the study of the order parameters in compounds at optimal doping. The most relevant results obtained in various systems were discussed in detail elsewhere \cite{daghero10,daghero11}. Here we will just try to give a general view of the open problems and thus of the present challenges to PCARS.
In the 1111 family, the main debated points about the results of PCARS measurements are the \emph{number} of the superconducting gaps and their possible anisotropy. Generally, the PCARS spectra in these compounds feature two clear symmetric peaks, as expected for a full superconducting gap (see for example fig.\ref{fig:symmetry}a and b). The fit of the spectra with a Blonder-Tinkham-Klapwijk (BTK) model \cite{BTK} gives a gap amplitude $\Delta_1 \leq \Delta_{BCS}$ -- the relevant gap ratio $2 \Delta_1/k_BT_c$ ranges from 2.2 in $\mathrm{LaFeAsO_{1-x}F_{x}}$ with $T_c= 28.6$ K \cite{gonnelli09b} to 3.69 in $\mathrm{SmFeAsO_{0.85}F_{0.15}}$ with $T_c=42$ K \cite{chen08b}. In the rather few cases where its temperature dependence can be followed up to $T_c$, it is fairly close to that expected within the BCS theory \cite{chen08b,daghero09b,dmitriev10}. Generally, the variation of this gap in a single sample is rather small, unless a large variation in the local $T_c$ is also detected \cite{gonnelli09a} (with perhaps the exception of $\mathrm{EuFeAsO_{1-x}F_x}$ \cite{dmitriev10}). This, and the fact that most of the PCARS measurements have been performed in polycrystals -- thus with random direction of current injection -- seems to indicate that the small gap has a fairly good spatial homogeneity and is likely to show a small anisotropy in the $k$ space.
Although a few spectra show only the small gap features and can indeed be well fitted by a single-gap BTK model, \cite{chen08b,yates08a,samuely09a,dmitriev10,miyakawa10}, in most cases additional features are also present in the spectra, in the form of more or less marked shoulders {\color{blue}{(see for example  Fig.\ref{fig:symmetry} a and b)}}, that prevent a single-gap BTK model from fitting the spectra in the whole energy range. The amplitude and the shape of these features can vary very much from contact to contact; interestingly enough, they seem to be weaker when the current is injected along the $c$ axis, as in the case of PCARS in oriented films of La-1111 and Sm-1111 \cite{naidyuk10}, while if the current is injected along the $ab$ plane (as in the only PCARS measurements carried out in 1111 in single crystals so far) they are clearly detected \cite{karpinski09}. Also their position (in energy) can show marked variations in different contacts. On this basis, some authors decided to neglect these structures in the fit, judging that they are not due to a superconducting gap \cite{chen08b,naidyuk10}. This led to the claim of a single, BCS gap in FeSC \cite{chen08b}. Others, including us, decided instead to treat these shoulders as if they were the hallmark of a second gap $\Delta_2$ \cite{daghero09b,gonnelli09a,samuely09a,miyakawa10} and thus fitted the PCARS spectra using a two-band BTK model (with two isotropic gaps), obtaining in many cases a very good fit in a wide energy range {\color{blue}{(as shown in Fig. \ref{fig:symmetry}a and b)}}.  The resulting second gap $\Delta_2$ has a very large ratio $2 \Delta_2 /k_B T_c$ generally between 6 and 10, and its amplitude can show considerable variations from one spectrum to another \cite{daghero09a,gonnelli09a,gonnelli09b}. This scatter can have extrinsic causes, i.e. the asymmetry of the PCARS curve for positive/negative bias \cite{naidyuk10}, the uncertainty in the normalization \cite{daghero11}, the broadening of the relevant structures which are not as neat as the ones associated with $\Delta_1$. However, intrinsic origins, such as a marked anisotropy of the large gap in $k$ space, cannot be excluded. ARPES measurements in $\mathrm{NdFeAsO_{0.9}F_{0.1}}$ \cite{kondo08} have shown a gap of about 15 meV (even larger than $\Delta_2$ measured by PCARS in the same compound \cite{samuely09a}) on the holelike pocket around the $\Gamma$ point; the data do not show a clear angular dependence of the amplitude in the $(k_x,k_y)$ plane but allow concluding that, if such a dependence is present, the in-plane anisotropy is no more than  20\% of the average value (i.e. of about 3 meV). In 1111 the calculated holelike FS is almost 2D, so if the opening of the large gap on this surface is a general feature of these compounds, its weight in the PCARS spectra \emph{is} expected to be minimum for $c$-axis current injection, as it happens for the $\sigma$-band gap in MgB$_2$ -- while the ubiquity of the small gap seems to require a larger band dispersion along $k_z$, i.e. a larger warping of the relevant FS sheet.
It follows from the above that PCARS measurements in 1111 compounds generally agree on the existence of one or more full gaps in these compounds, with neither zeros nor nodes. However, the emergence of nodes in 1111 compounds with low $T_c$ has been shown to be possible as a consequence of the decrease in the pnictogen height \cite{kuroki09}. To the best of our knowledge, no attempts to perform PCARS measurements in compounds that could fall in this region have been made yet. In this concern, however, it is worth recalling that the claims of nodal superconductivity in $\mathrm{LaFePO}$ based on penetration-depth measurements \cite{fletcher09,hicks09} have been recently questioned \cite{bang12}.

In compounds of the 122 family, the multiple-gap scenario is universally accepted, also thanks to ARPES measurements that have unambiguously shown gaps with different amplitude residing on different Fermi surface sheets (for a review see Ref. \cite{richard11}). Favored by the early availability of large single crystals, directional PCARS measurements were carried out very soon in $\mathrm{(Ba,K)Fe_2As_2}$ \cite{samuely09b,szabo09,lu10}. {\color{blue}{An example of PCARS spectrum in a $\mathrm{(Ba,K)Fe_2As_2}$ crystal, taken from Ref.\cite{szabo09} and obtained with in-plane current injection is shown in fig.\ref{fig:symmetry}d.}} As shown elsewhere \cite{daghero11} $ab$-plane \cite{samuely09b,szabo09} and $c$-axis \cite{lu10} spectra obtained in this compound are perfectly consistent and support \emph{isotropic} gaps $\Delta_1$ and $\Delta_2$ (see the 2D BTK fit of fig.\ref{fig:symmetry}d) with gap ratios $2\Delta_1/k_BT_c\simeq 2.6$ and $2\Delta_2/k_BT_c\simeq 9$, although the weight of the latter in the conductance seems to be suppressed for $c$-axis current injection. $\mathrm{(Ba,K)Fe_2As_2}$ has also been the object of the only PCARS study as a function of doping \cite{zhang10}. The authors claim a universal and isotropic $2\Delta/k_B T_c$ ratio of about 3.1 in the whole doping range and also in the electron-doped compound $\mathrm{SrFe_{1.74}Co_{0.26}As_2}$. Unfortunately, they use a single-gap approach so that their fit is not satisfactory in the whole voltage range, and possibly interesting details (such as zero-bias cusps) are completely disregarded.

At present, the strongly debated point is the presence or absence of (accidental or symmetry-imposed) nodes in some of the 122 systems.
In $\mathrm{KFe_2As_2}$, for instance, a nodal symmetry such as $d$-wave has been proposed on the basis of resistivity and thermal conductivity measurements \cite{dong10,reid12}; in $\mathrm{BaFe_2(As,P)_2}$ lines nodes have been suggested to explain the results of penetration depth, thermal transport and NMR measurements \cite{hashimoto10,nakai10}. No PCARS measurements have been carried out so far in these compounds to verify this possibility.
According to theoretical calculations by Suzuki et al. \cite{suzuki11} accidental node lines with a complex 3D geometry within a general $s \pm$ symmetry can emerge (in a otherwise fully gapped compound) on the region of the outer hole Fermi surface around the Z point, when this portion of the FS, having a strong $3Z^2-R^2$ ($Z^2$) orbital character becomes large by isovalent doping. Unlike in 1111, the emergence of nodes would not require low $T_c$ because the $Z^2$ orbitals do not play an important role in the spin-fluctuation-mediated superconductivity. Recently, however, the interpretation of the experimental data in $\mathrm{BaFe_2(As,P)_2}$ as suggesting nodal lines in the gap has been seriously questioned \cite{bang12}.

Claims of accidental nodes within a general $s\pm$ symmetry have also been made in $\mathrm{Ba(Fe,Co)_2As_2}$ based on Raman spectroscopy \cite{muschler09} and directional thermal conductivity \cite{reid10}. Consistently, a residual linear term in the specific heat was observed \cite{gofryk10,gofryk11,hardy10}, which is minimum at optimal doping. These results were in turn explained theoretically as being due to gap nodes or deep gap minima that form elliptical loops on the outer electron FS centered at the intersection of the FS with the $\Gamma-M$ line.
In this case, a comparison with PCARS measurements is possible, at least around optimal doping. PCARS measurements carried out in $\mathrm{Ba(Fe_{1-x}Co_x)_2As_2}$ single crystals with $x=0.07$ \cite{samuely09b} and $x=0.10$ \cite{tortello10} show no detectable signs of nodes, and the relevant spectra can be well fitted to a 2D BTK model with one \cite{samuely09b} or two gaps \cite{tortello10}. In the latter case, a very small anisotropy of the two gaps is observed, since $\Delta_1=4.4 \pm 0.6$ meV and $\Delta_2=9.9 \pm 1.2$ meV in the $ab$ plane, while $\Delta_1=4.1 \pm 0.4$ meV and $\Delta_2=9.2 \pm 1.0$ meV along the $c$ axis. {\color{blue}{Similar gap amplitudes are given by the 2D-BTK fit (with isotropic gaps) of PCARS spectra in $\mathrm{Ba(Fe_{1-x}Co_x)_2As_2}$ films with nominal $x=0.08$ (see fig.\ref{fig:symmetry}c)}}. However, if a more sophisticated 3D-BTK model for the fit of the PCARS spectra is used -- taking into account the geometry of the Fermi surface -- an \emph{excess conductance at zero bias} is found in the experimental spectra, {\color{blue}{especially if the current is injected along the $c$ axis \cite{gonnelli12b}}}, whose possible origin will be discussed in Sect. \ref{sect:3DBTK}.

There are few examples of PCARS measurements that clearly indicate anisotropic or nodal gaps. In $\mathrm{Ca(Fe_{1-x}Co_x)_2As_2}$ single crystals with $x=0.060 \pm 0.005$, \emph{all} the spectra show a clear zero-bias maximum that is strongly suggestive of an anisotropic or nodal gap. {\color{blue}{An example of these spectra is shown in fig.\ref{fig:symmetry}d, together with its 2D-BTK fit obtained by using an isotropic large gap $\Delta_2$ and a fully anisotropic small gap $\Delta_1(\theta)=\Delta_1 \cos^4[2(\theta-\alpha)]$. The fit is also possible if a 2D-BTK model with a small gap in $d$-wave symmetry is used, i.e. $\Delta_1(\theta)=\Delta_1 \cos[2(\theta-\alpha)]$, because the presence of broadening effects makes the distinction between these two symmetries impossible \cite{daghero11}. However, a $d$-wave symmetry is not expected for this material; and moreover, any in-plane anisotropy is likely not to fully reflect the probable real structure of the gap.}} Therefore, we made a step forward and used for the fit the aforementioned 3D-BTK model (described in the Sect. \ref{sect:3DBTK}) that makes use of analytic surfaces to model the real Fermi surface. According to DFT calculations, at $x=0.060$ the outer holelike sheet undergoes a topological transition \cite{gonnelli12} and splits into separate closed pockets centered at the Z points. The situation can be viewed as an extreme evolution of that discussed by Suzuki et al. \cite{suzuki11}, and so should be also the gap structure. We thus assumed the gap on the electron pockets to have vertical node lines (with change of sign) \cite{gonnelli12} or lines of zeros (with no sign change) \cite{gonnelli12b} intersecting each other along the $\Gamma$-Z line. The fit is possible in both cases since, as in the 2D case, the large broadening prevents a distinction between the two symmetries \cite{daghero11}.
Very recently, PCARS measurements in single crystals of the optimally electron doped compound $\mathrm{BaFe_{1.9}Ni_{0.1}As_2}$ have been reported \cite{wang12}. The $c$-axis spectra show clear double-gap structures, i.e. a broad maximum centered at zero bias plus shoulders at about 7 meV, and were very well fitted by a two-band BTK model with one isotropic gap $\Delta_h =4.2-4.4$ meV on the hole FS and a gap $\Delta_e$ on the electron FS, with an in-plane anisotropy (i.e. lines of minima parallel to the $c$ axis) and maximum amplitude of 9.8-10.5 meV.
Once again, in both this case and that of $\mathrm{Ca(Fe_{1-x}Co_x)_2As_2}$ (as well as in all the other compounds where the emergence of nodes is possible) it would be interesting to see what happens at other doping contents; possibly, exploring the phase diagram of the compound by PCARS might help tracking the entire evolution from full gaps to nodal gaps as predicted theoretically in 122 systems \cite{suzuki11}.

\subsection{The pairing mechanism} \label{subsect:pairing}
\begin{figure*}[ht]
\includegraphics[width=0.9\textwidth]{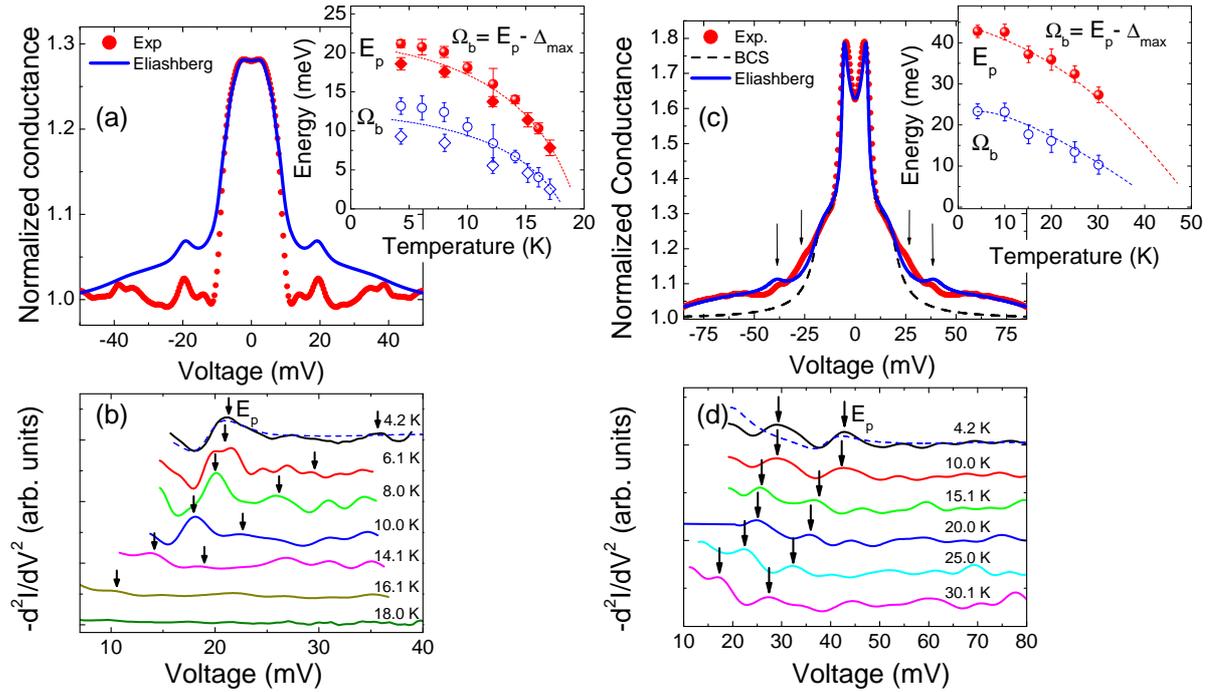}
\caption{(a): normalized AR conductance curve (symbols) measured on Ba(Fe$_{0.9}$Co$_{0.1}$)$_2$As$_2$ compared to the theoretical one (line) obtained from Eliashberg (energy-dependent gaps) and BTK calculations. (b): temperature dependence of the $-d^2 I / d V^2$ curves (solid lines) showing the displacement of the bosonic structures. Dash line is the theoretical $-d^2 I / d V^2$ obtained from the line in (a). The energy of the peak E$_p$(T) and the corresponding characteristic boson energy $\Omega_b$(T) are shown in the inset to panel (a); lines are only guide to the eye. (c) and (d): the same as in (a) and (b) but for SmFeAsO$_{0.8}$F$_{0.2}$. Dash line in (c) is the theoretical AR spectrum calculated with BCS (energy-independent) gaps.}\label{fig:EPI}
\end{figure*}

It is well known that, owing to the proximity of superconductivity to a static magnetic order, a pairing mechanism mediated by SFs has been soon accepted as the most convincing one for FeSC. The existence of a spin resonance that appears below $T_c$ evidenced by studies of the spin dynamics \cite{lumsden10} strongly supports the role of SFs in the formation of a superconducting condensate. PCARS offers the unique opportunity to \emph{directly} detect the coupling between electrons and SFs. It is true that the PCS in the \emph{normal} state is the most traditional way to extract the spectrum of excitations that give rise to electron scattering \cite{yanson74,naidyuklibro,yanson97}; but in the case of FeSC the normal state at low temperature is not accessible due to the very high upper critical fields and the $T_c$ is too high for PCS measurements in the normal state \emph{above} $T_c$ to ensure the sufficient energy resolution. Fortunately, in the strong-coupling regime information on the spectrum of the mediating boson can be obtained also from PCARS in the superconducting state.
Indeed, in the Eliashberg theory \cite{eliashberg60} the order parameter is a complex function of energy $\Delta(E)=\Re \Delta(E) +i\Im\Delta(E)$ whose imaginary part accounts for the finite lifetime of Cooper pairs and retains information about the electron-boson spectrum (or Eliashberg function) $\alpha^2F(\omega)$ . As a consequence, also the superconducting DOS $N(E)=\Re\left[E/(E^{2}-\Delta^{2})\right]$ shows deviations from the BCS one due to the electron-boson interaction \cite{wolf85}. If one performs \emph{tunnel spectroscopy} which, at low temperature, is sensitive to the DOS, a peak in $\alpha^2 F(\omega)$ at an energy $\Omega_0$ will give rise to a peak in the second derivative of the I-V curve ($-d^2 I/dV^2$) at an energy $E_p=\Omega_0+\Delta_{max}$ where $\Delta_{max}$ is the larger gap. In a single-band superconductor, the shape of the entire $\alpha^2 F(\omega)$ can be extracted from the tunneling curve \cite{naidyuklibro}. Away from the pure tunneling regime, i.e. when the potential barrier at the interface decreases and one goes toward the Andreev regime, the conductance is no longer proportional to the DOS but the electron-boson features persist, although the peak in $-d^2 I/dV^2$ decreases in amplitude (this can be understood by inserting $\Delta(E)$ into the BTK model \cite{tortello10,daghero10,daghero11}). Strictly speaking, in a multiband system it is not possible to extract the $\alpha^2 F(\omega)$ form the PCARS spectrum \cite{dolgov03}, but one can try to understand what is the electron-boson spectrum that, once inserted into the Eliashberg theory, allows reproducing the observed features in the conductance and its derivative.
Of course, all the procedure relies on the fact that, as shown in detail elsewhere \cite{ummarino09}, Eliashberg theory works -- at least as a phenomenological description -- in Fe-based superconductors.
So far, we have studied the electron-boson coupling in $\mathrm{Ba(Fe_{0.92}Co_{0.08})_2As_2}$ \cite{tortello10,daghero11} (see fig.\ref{fig:EPI} a and b) and $\mathrm{SmFeAsO_{0.8}F_{0.2}}$ \cite{daghero11} (see fig.\ref{fig:EPI} c and d). In both cases, we started from PCARS spectra that showed particularly clear additional features, also detectable as peaks in the second derivative $-d^2 I/dV^2$ (Fig. \ref{fig:EPI} b and d) and we showed that the experimental derivatives of the conductance $-d^2 I/dV^2$ but also the conductance itself can be reproduced starting from a Lorentzian spectrum {\color{blue}{of the bosonic excitations}} (that mimics the spin-fluctuation one calculated in the superconducting state \cite{nagai11}) peaked at an energy $\Omega_0$ that: i) in  $\mathrm{Ba(Fe,Co)_2As_2}$ coincides, within the uncertainty, with the energy of the spin resonance measured by inelastic neutron scattering \cite{inosov10}, and ii) in $\mathrm{SmFeAs(O,F)}$ fulfills the empirical relationship $\Omega_0=4.65k_B T_c$ \cite{paglione10}. A spectrum similar to that calculated for SFs in the normal state \cite{bose03} instead does not allow reproducing the electron-boson features unless the energy cutoff is unphysically low \cite{daghero11}. On increasing the temperature $T$, the boson energy $\Omega(T)=E_p(T)-\Delta_2(T)$ also decreases, following the trend shown in the insets to Fig. \ref{fig:EPI} a and b. This clearly indicates the non-phononic origin of the bosonic mode. The conclusion is that a mediating boson with a spectrum similar to that of SFs, and peaked at the energy of the spin resonance in the superconducting state, allows reproducing the additional structures in the PCARS spectra of two very different FeSC compounds. It is worth noticing that PCARS detects features related to the \emph{interaction of electrons with elementary excitations} in the material (and not only to the excitation itself) as instead other spectroscopic techniques do, and in this sense it is a more direct tool for the determination of the electron coupling and provides a more convincing demonstration of the spin-fluctuation pairing in FeSC. Notice that, once this is established, the determination of the SF characteristic energy from the $-d^2I/dV^2$ curves is straightforward. For example, Wang et al. \cite{wang12} have recently used this procedure to extract the characteristic energy of the mediating boson in optimally-doped $\mathrm{Ba(Fe_{1.9}Ni_{0.1})_2As_2}$. It would be thus very useful, and also rather simple, to study for example the evolution of the boson frequency of a given system with doping. This piece of information could be essential for a thorough study of the spin dynamics in the whole phase diagram and in particular in the in underdoped samples, to better understand the crossover from static magnetic order to the spin fluctuation regime.

\subsection{Coexistence of magnetism and superconductivity}\label{subsect:m_and_SC}
\begin{figure}[ht]
\includegraphics[width=0.9\columnwidth]{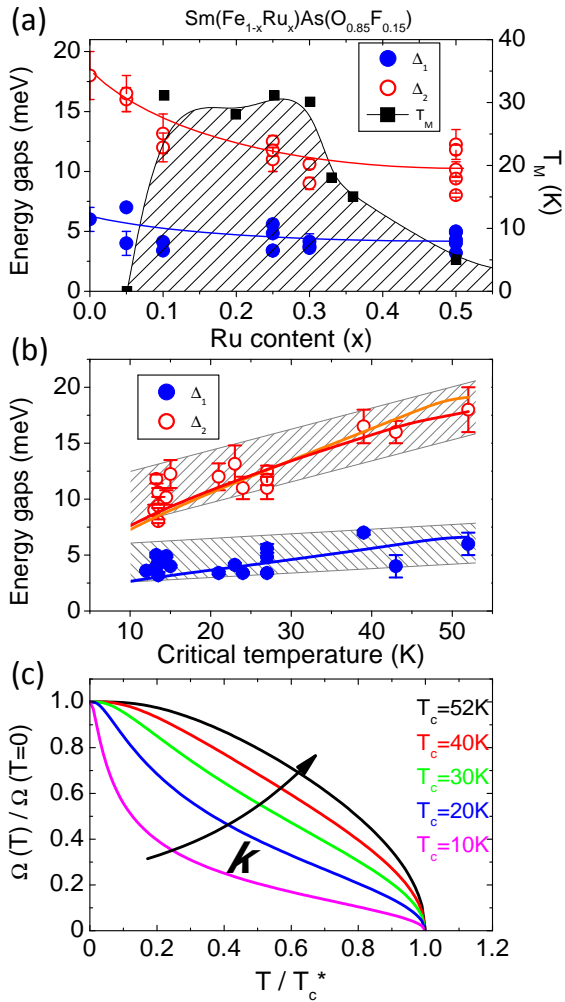}
\caption{(a) Temperature dependence of the magnetic transition temperature T$_M$ (squares) and gaps (circles) on the Ru content ($x$) in Sm(Fe$_{1-x}$Ru$_x$)As(O$_{0.85}$F$_{0.15}$). (b) Temperature dependence of the gaps shown in (a) but plotted as a function of the local T$_c$ of the contacts. Lines are calculations within the three-band $s \pm$-wave Eliashberg theory. (c) Temperature-dependent part of the superfluid density used in the Eliashberg calculations shown in (b).}\label{fig:magnetic}
\end{figure}
What is the exact nature of the region of coexistence of magnetic order and superconductivity in the ``underdoped'' part of the phase diagram of most FeSC is still not clear. In the hole-doped system $\mathrm{(Ba,K)Fe_2 As_2}$ a spatial phase separation seems to occur, while a true microscopic coexistence of magnetic and superconducting order parameters has been claimed in the electron-doped $\mathrm{Ba(Fe,Co)_2As_2}$ \cite{julien09,lumsden10}. PCARS investigations in this region might help clarifying the effect of the magnetic order on the spectra and possibly on the superconducting gaps. Indeed, ``anomalous'' features such as a zero-bias peak that disappears \emph{above} the superconducting critical temperature have been sometimes detected in $\mathrm{FeSe_{0.45}Te_{0.55}}$ \cite{park10} and $\mathrm{Ba(Fe_{0.92}Co_{0.08})_2 As_2}$ \cite{sheet10}. In the first case, Park et al. interestingly note the similarity between this peak and that observed in Cd-doped $\mathrm{CeCoIn_5}$, that persists above the bulk $T_c$ and up to the N\'{e}el temperature $T_N$. A mechanism of quasiparticle scattering off an antiferromagnetic order is proposed to explain this feature \cite{bobkova05}; this might be very likely to occur in the region of coexistence even though in the specific case the samples should be at optimal doping. PCARS measurements in the underdoped region of the phase diagram, i.e. at doping contents where a magnetic order still survives, are necessary in the near future to address this point.
Recently, $\mu$SR and NQR  measurements have shown the recovery of a short-range static magnetic order and the concomitant degradation of the superconducting $T_c$ as a result of isoelectronic substitution of Fe with Ru in optimally F-doped $\mathrm{SmFeAsO_{0.85}F_{0.15}}$ \cite{sanna11}. The magnetic and superconducting order parameters have been found to coexist within nanometer-size domains in the FeAs layers and to eventually disappear around a common critical threshold in the Ru content (60\%).  The dependence of the magnetic transition temperature $T_M$  on the Ru content ($x$ in $\mathrm{Sm(Fe_{1-x}Ru_x)AsO_{0.85}F_{0.15}}$) is shown in Fig. \ref{fig:magnetic}a (solid squares). We performed PCARS measurements in these samples up to $x=0.50$. It is immediately clear from the raw experimental data that: i) all the spectra show more or less marked double-gap features, i.e. the two-gap character is retained up to the highest Ru doping; ii) none of the spectra display zero-bias peaks that may suggest the emergence of nodes, i.e. the symmetry and also the structure of the order parameters presumably remain the same as in $\mathrm{SmFeAsO_{0.85}F_{0.15}}$; iii) despite the very wide doping range, the width (in energy) of the Andreev reflection structures (and thus, roughly speaking, the amplitude of the gaps) does not change very much \cite{daghero12}. Given the above, the amplitudes of the two (supposed isotropic) gaps were extracted as usual by fitting the spectra with a two-band 2D-BTK model with adjustable relative weight of the two partial conductances associated to $\Delta_1$ and $\Delta_2$. The trend of the gaps as a function of the Ru content is shown in Fig.\ref{fig:magnetic}a (open and full circles). A vertical spread of the data is evident, that accounts for their variation over different contacts. Since also the local $T_c$ is found to change, a plot of the gaps as a function of the local $T_c$ removes the degeneracy, and clearly shows an almost linear trend of both $\Delta_1$  and $\Delta_12$ vs $T_c$ \cite{daghero12} (see Fig.\ref{fig:magnetic}b). Despite this correlation, the gap ratios $2\Delta_i /k_B T_c$ do \emph{not} remain constant but increase, first slowly and then more steeply, on decreasing $T_c$. The trend is perfectly superimposed to the general one obtained by plotting the gap ratios of different compounds, measured by PCARS or ARPES, as a function of the relevant $T_c$ \cite{daghero11}. This trend, which thus seems to be universal in Fe-based compounds, looks in contradiction with the fact that the characteristic boson frequency $\Omega_0$ (or, rather, the energy of the spin resonance measured by inelastic neutron scattering) linearly \emph{decreases} on decreasing $T_c$ \cite{inosov10,paglione10}. Even with extremely big values of the coupling strengths, it is not possible to obtain such large gaps within the Eliashberg theory with such a small $T_c$ using the experimental values of the spin-fluctuation frequency. The key to solve the puzzle is the \emph{feedback effect} of the condensate on the  mediating boson, which is due to the fact that both of them arise from the same electronic excitations. The temperature dependence of the boson frequency is indeed similar to that of the superfluid density \cite{ummarino11}, i.e. $\Omega_0(T)=\Omega_0 \eta(T)$ where $\eta(T)$ is the temperature dependent part of the superfluid density: $\rho(T)=\rho_0 \eta(T)$. Penetration-depth measurements in underdoped $\mathrm{Ba(Fe,Co)_2As_2}$ samples showed, as expected, a suppression in $\rho_0$ due to the competition of magnetic and superconducting order \cite{prozorov11}, but also a change in $\eta(T)$, that deviates from the BCS trend and shows a convex shape and a positive curvature at $T> T_c/2$. A general functional form that can mimic this evolution in the $\Omega(T)$ curve is $\Omega_0(T)= \Omega_0 \tanh{\left(1.76 k \sqrt{T_c^* /T-1}\right)}$ being $T_c^*$ the critical temperature without feedback effect. The parameter $k$ can be varied until the experimental $T_c$ is reproduced within the Eliashberg theory ($T_c \ll T_c^*$); then a test of self consistency can be made by calculating the superfluid density and verifying that it approximates the analytical $\Omega_0(T)$ curve. In the sample with no Ru,  $k=0.6192$ and the relevant $\Omega_0(T)$ is shown in Figure \ref{fig:magnetic}c as a black line. Upon Ru doping, the decrease in $T_c$ requires a decrease in $k$ that means a more and more marked change in the temperature dependence of the boson frequency, as shown in Fig. \ref{fig:magnetic}c \cite{daghero12}.  In conclusion, in the case of $\mathrm{Sm(Fe_{1-x}Ru_x)AsO_{0.85}F_{0.15}}$ PCARS spectra do not show any feature that can be directly related to the onset of short-range magnetic order competing with superconductivity. However, this order (observed by $\mu$SR  and NQR \cite{sanna11}) \emph{does} have an effect that consist in an anomalous suppression in the superfluid density (and thus in the characteristic spin-fluctuation energy) that, in turn, allows very large gaps to be compatible with low $T_c$ values. The challenge now is, first, to compare these results with direct measurements of superfluid density in the same samples, and then to understand if, and to what extent, this piece of information can be used to explain the apparently universal increase in the gap ratio in samples with low $T_c$ \cite{daghero11,daghero12}.

\section{The 3D BTK model and the zero-bias anomalies}\label{sect:3DBTK}
In the previous section we have seen that almost all the PCARS measurements in Fe-based compounds have been so far interpreted and analysed in the framework of the 2D-BTK model described in the seminal papers of S. Kashiwaya, Y. Tanaka et al. \cite{kashiwaya96}. To account for the multiband nature of these superconductors, \emph{effective} two-band models have been used by simply expressing the total normalized conductance as the weighted sum of the partial contributions of each band \cite{daghero10,daghero11}. The fitting parameters are already seven: $\Delta$ (maximum amplitude of the gap), $\Gamma$ (broadening parameter \cite{plecenik94}) and $Z$ (dimensionless height of the barrier) for each band, plus the weight $w$ of the contribution of one band with respect to the other. Of course, the use of more than two bands would make the fit meaningless due to the large number of free parameters. However, one of the basic assumptions of the 2D-BTK model is that the Fermi surface (FS) is spherical (circular) in both the N and the S side of the junction.
This model can properly describe the isotropic superconductors or the layered ones, like cuprates, provided that the current is injected along the $ab$ plane. The problem of how to express the $c$-axis conductance in layered superconductors, or more generally, the conductance in multiband superconductors with complex shapes of the FS sheets has to be solved with a proper generalization of the model. This generalization that, of course, can be rather demanding from the numerical point of view has been recently introduced by us \cite{daghero10,daghero11}. A complete description of the resulting ``3D-BTK model'' is beyond the scope and the available space of the present review but all the details can be found in the recent literature \cite{daghero11,daghero12b}.

In the 3D-BTK model the general expression of the normalized conductance of a point-contact NS junction at $T=0$ is given by:
\begin{equation}\label{eq:G_FS}
\langle
G(E)\rangle_{I\parallel\mathrm{\mathbf{n}}}=\frac{\sum_{i}\langle\sigma_{i\mathrm{\mathbf{k}}n}(E) \tau_{\mathit{i}\mathrm{\mathbf{k}},n} \frac{v_{\mathit{i}\mathrm{\mathbf{k}},n}}{v_{\mathit{i}\mathrm{\mathbf{k}}}}\rangle_{FS_{i}}}{\sum_{i}\langle\tau_{\mathit{i}\mathrm{\mathbf{k}},n} \frac{v_{\mathit{i}\mathrm{\mathbf{k}},n}}{v_{\mathit{i}\mathrm{\mathbf{k}}}}\rangle_{FS_{i}}}
\end{equation}
In this equation, $i$ is the band index, the brackets indicate the average over the $i$-th Fermi surface sheet, the subscript $n$ refers to the direction of current injection and $v_{\mathit{i}\mathrm{\mathbf{k}},n}=\mathbf{v}_{\mathit{i}\mathrm{\mathbf{k}}} \cdot \mathbf{n}$ is the projection of the Fermi velocity on the $i$-th band (that is thus perpendicular to the $i$-th FS sheet) along the direction of the unit vector $\mathbf{n}$ parallel to the injected current. The normal-state barrier transparency is defined as
\begin{equation}\label{eq:tau2}
\tau_{\mathit{i}\mathrm{\mathbf{k}},n}=\frac{4v_{\mathit{i}\mathrm{\mathbf{k}},n}v_{N,n}}{(v_{\mathit{i}\mathrm{\mathbf{k}},n}+v_{N,n})^{2}+4Z^{2}v_{N}^{2}}
\end{equation}
where $v_{N,n}= \mathbf{v}_N \cdot \mathbf{n}$, $\mathbf{v}_N$ being the Fermi velocity in the normal material supposed constant in magnitude (spherical FS).
The quantity $\sigma_{i\mathrm{\mathbf{k}}n}(E)$ is the superconducting-state relative barrier transparency expressed by the following equation:
\begin{equation}
\label{eq:sigma_anis}
\sigma_{i\mathrm{\mathbf{k}}n}(E)=\frac{1+\tau_{\mathit{i}\mathrm{\mathbf{k}},n}|\gamma_{+}(E)|^{2}+(\tau_{\mathit{i}\mathrm{\mathbf{k}},n}-1)|\gamma_{+}(E)\gamma_{-}(E)|^{2}}{|1+(\tau_{\mathit{i}\mathrm{\mathbf{k}},n}-1)\gamma_{+}(E)\gamma_{-}(E)\exp(i\varphi_{d})|^{2}}
\end{equation}
where
\begin{equation}
\gamma_{\pm}(E)=\frac{(E+i\Gamma)-\sqrt{(E+i\Gamma)^{2}-|\Delta_{\mathit{i}\mathbf{k}}^{\pm}|^{2}}}{|\Delta_{\mathit{i}\mathbf{k}}^{\pm}|} \label{eq:gammapm}
\end{equation}
and
\begin{equation}
\varphi_{d}=-i\ln\left[\frac{\Delta_{\mathit{i}\mathbf{k}}^{+}/\left|\Delta_{\mathit{i}\mathbf{k}}^{+}\right|}{\Delta_{\mathit{i}\mathbf{k}}^{-}/\left|\Delta_{\mathit{i}\mathbf{k}}^{-}\right|}\right] \label{eq:varphi}
\end{equation}
is the phase difference seen by the holelike quasiparticles (HLQ) with respect to the electronlike ones (ELQ), being $\Delta_{\mathit{i}\mathbf{k}}^{+}$ and $\Delta_{\mathit{i}\mathbf{k}}^{-}$ the different (in phase and/or in magnitude) $\mathbf{k}$-dependent order parameters of the $i$-th band felt by the ELQ  and by the HLQ, respectively. By looking at the previous equations it is clear that the complete knowledge of the $\mathbf{k}$ dependence of the FS sheets (i.e. their geometry in the reciprocal space) and of the order parameters (i.e. their symmetry), under some simplifying conditions, can give us all the ingredients required to calculate the normalized conductance. Indeed if we suppose that the points at the Fermi energy are close to the top or the bottom of parabolic-like bands (where the effective mass approximation holds) the Fermi velocity at any $\mathbf{k}$ point can be uniquely expressed as a function of the constant effective mass and of the FS shape and dimensions \cite{daghero11}.

It is rather instructive to briefly describe two limiting cases of the general equation for the normalized conductance, in the restrictive conditions of spherical Fermi surfaces and isotropic order parameters. When the barrier at the NS interface has a negligible transparency (i.e. $Z\rightarrow \infty$), i.e. in a tunnel junction, eq. \ref{eq:G_FS} reduces to a weighted average of the relative superconducting-state transparencies of the bands $\sigma_{i}(E)$ where the weights can be expressed as $\langle N_{i\mathrm{\mathbf{k}}} v_{\mathit{i}\mathrm{\mathbf{k}},n}^2\rangle_{FS_{i}} $ \cite{mazin99,daghero12b}, $N_{i\mathrm{\mathbf{k}}}$ being the normal density of states (DOS) of the \emph{i}-th band at the Fermi energy and wave vector $\textbf{k}$ in the superconductor.
When instead the barrier at the NS interface is completely transparent (i.e. $Z$ = 0), the total normalized conductance is again a weighted average of the $\sigma_{i}(E)$ but the weights can be now expressed as $\langle N_{i\mathrm{\mathbf{k}}} v_{\mathit{i}\mathrm{\mathbf{k}},n}\rangle_{FS_{i}}$ \cite{mazin99,daghero12b}. Since the normal DOS is always proportional to the reciprocal of the Fermi velocity, it turns out that the previous average on the FS returns the area of the projection of the $i$-th FS sheet on a plane perpendicular to $\mathbf{n}$.
The implications of this result is that, in the case of ballistic point contacts on normal metals (NN' junctions with $Z$=0) the conductance is \emph{not} expected to contain any information on an energy-dependent DOS.

\begin{figure*}[ht]
\includegraphics[width=0.8\textwidth]{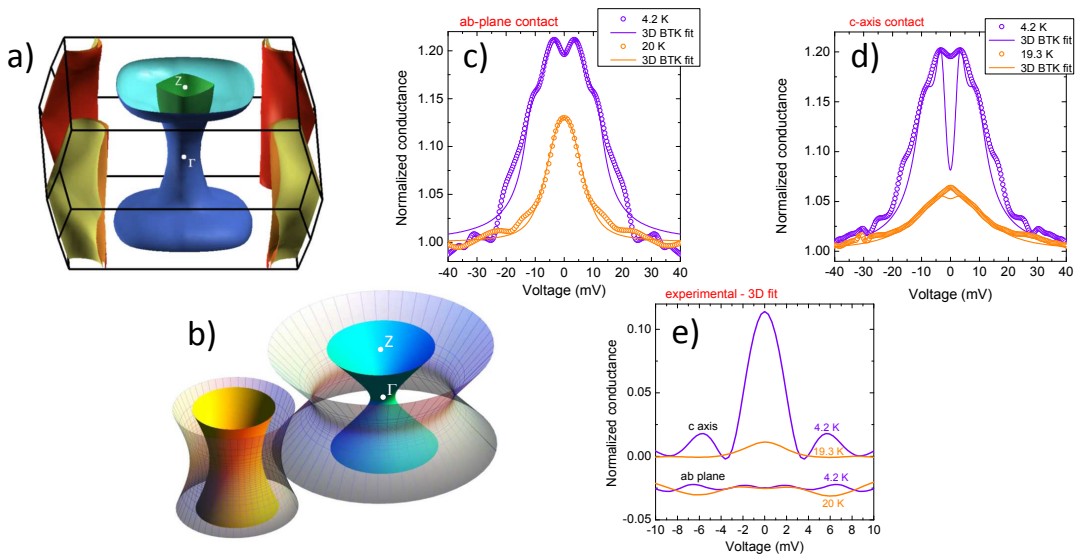}
\caption{(a): Fermi surface of Ba(Fe$_{0.9}$Co$_{0.1}$)$_2$As$_2$. (b): the model FS used in the 3D-BTK model to fit the experimental conductance curves. Solid surfaces represent the FS sheets, while gridded surfaces represent the amplitude of the corresponding energy gap. (c): normalized PCARS conductance curves (symbols) measured at two different temperatures in an $ab$-plane contact on Ba(Fe$_{0.92}$Co$_{0.08}$)$_2$As$_2$ together with their relevant 3D-BTK fitting curves (lines). (d): the same as in (c) but for a $c$-axis contact. (e): differences between experimental and theoretical curves shown in (c) and (d). Curves referring to the $ab$-plane contact are shifted downward for clarity.}
\label{fig:3DBTK}
\end{figure*}

Figure \ref{fig:3DBTK} illustrates the various steps we adopted for fitting PCARS conductance curves by using the 3D-BTK model in a real case.  We will refer here to directional PCARS experiment we performed on high-quality single crystals of $\mathrm{Ba (Fe_{1-x}Co_{x})_2 As_2}$ with nominal $x=0.1$ (bulk $T_c=24.5$ K). The spectra, previously fitted by using a two-band 2D BTK model with two isotropic gaps, are already present in literature  \cite{tortello10}. The first step of the 3D BTK approach consists in calculating the FS of the Fe-based superconductor at the proper doping within the Density-Functional Theory (DFT) \cite{daghero11}.
The results are shown in Fig. \ref{fig:3DBTK}a. As shown in several earlier papers \cite{singh08b, mazin09,vilmercati09} the FS features two holelike warped cylinders around the $\Gamma-Z$ axis (with the warping of the outer one much more marked) and two electronlike warped cylinders with the characteristic ``elliptical'' cross-section that depends on $k_z$. The next step consists in modeling the actual FS by means of two analytical surfaces whose relative dimensions, shapes and curvatures are as close as possible to the ones shown in Fig. \ref{fig:3DBTK}a. As already mentioned the number of model FS sheets has to be restricted to two in order to keep the number of free parameters of the model at a reasonable value  (6 in this case). Fig. \ref{fig:3DBTK}b shows this model FS that is made up of two hyperboloids of revolution, one for the strongly-warped outer holelike FS sheet (the inner one is neglected) and one for the two almost-degenerate electronlike sheets (matt surfaces). As already pointed out, in order to calculate the PCARS spectra it is also necessary to assume the $\mathbf{k}$ dependence of the two order parameters on the FS sheets or, in other words, to fix their symmetry. Since a number of experimental results reported in literature (including ARPES \cite{terashima09} and our previous 2D-BTK analysis of the same data \cite{tortello10}) indicate two isotropic gaps for this compound, and for the sake of comparison with the 2D analysis, we assume here two isotropic gaps, $\Delta_1$ on the holelike FS and $\Delta_2$ on the electrolike FS (gridded surfaces in Fig. \ref{fig:3DBTK}b). It is now possible to use eq. \ref{eq:G_FS} (or better its convolution with the Fermi function at finite $T$) for fitting the experimental conductance curves. Four examples of this fitting procedure are shown in Fig. \ref{fig:3DBTK}c and d. In Fig. \ref{fig:3DBTK}c the normalized conductance at $T=4.2$ K and 20 K (symbols) measured by injecting the current along the $ab$ plane of the crystal ($ab$-plane ``soft'' point contact) is compared to the results of the 3D BTK fit for $I\parallel ab$ (solid lines). All the six parameters of the fit ($\Delta_{1,2}$,  $\Gamma_{1,2}$ and $Z_{1,2}$) are left adjustable, while it is worthwhile to underline that the weight $w$ of one band with respect to the other is no more a parameter since it is here uniquely determined by the FS shape, the direction of current injection and the $Z$ values. The results of the 3D fit are quite good for $eV < 15$ meV and the resulting values of the parameters (reported in the caption of Fig. 2) are very similar to the values already obtained by the 2D BTK fit \cite{tortello10}. This fact strongly supports the obvious conclusion that the 2D BTK fit returns the correct values of the parameters when most of the surface of FS sheets is almost perpendicular to the direction of current injection. The conductances at $eV > 15$ meV are largely dominated by structures that, as discussed in the previous section, have been associated to the effect of a strong electron-boson coupling to a spin-fluctuation spectrum with a characteristic energy of about 12 meV \cite{tortello10}. Of course these structures cannot be reproduced here where the explicit energy dependence of the order parameters is not taken into account. Fig. \ref{fig:3DBTK}d shows the results of the same procedure applied to ``soft'' point contacts on the same crystals where the current was injected along the $c$ axis. In this case it is quite evident that the theoretical curves systematically fail to fit the low-bias region of the spectra and show a zero-bias dip which is much deeper than the experimental one. The fit cannot be improved even if both $Z_1$ and $Z_2$ are taken close to zero. This is due to the so called $Z$-enhancing effect \cite{daghero11} that occurs when most of the surface of the relevant FS sheet is almost parallel to the direction of current injection (as in 2D-like FSs along the $c$ axis). In addition we have also noticed that the use of more complex symmetries for the gap on the holelike FS with horizontal and/or vertical nodes does not remove the zero-bias anomaly.  {\color{blue}{Neither does so the use of a different analytical model for the holonic FS; for example, the latter can be modelled as a pair of cup-shaped pockets connected by a cylinder (which is somewhat more appropriate than a single hyperboloid): indeed, the source of the big zero-bias dip is the \emph{electronic} FS and not the holonic one}}. One could argue that this apparent failure is related to a mistake in the formulation or application of the 3D BTK model. In order to rule out this possibility we recently re-analyzed the directional PCARS spectra we obtained in single crystals of MgB$_2$ and CaC$_6$ \cite{gonnelli02c,gonnelli08} by using the 3D-BTK model and exactly the same procedure presented here for fitting the conductance curves. In both cases the results are very good and an almost perfect agreement is obtained with the experimental PCARS conductances both along $ab$ plane and in $c$-axis direction by using the same gap values previously obtained within the 2D BTK fit \cite{daghero12b}. Since in both cases parts of the FS have a strong 2D-like character these results make us confident that the 3D model correctly describes the PCARS physics of materials with complex FSs.

{\color{blue}{One might still object that all the BTK models mentioned so far (including the 3D one) are formulated in the hypothesis of ballistic conduction through the point contact, that means in the hypothesis that the contact radius is much smaller than the electronic mean free path. However, owing to the high residual resistivity of FeSC and their consequently small electronic mean free path, this condition is likely not to be fulfilled in these materials. While a Maxwell regime can be certainly excluded based on the simple fact that the PCARS spectra of these contacts \emph{do} show spectroscopic features (and instead they do do not present features traditionally associated to heating effects \cite{daghero10,sheet04}), an intermediate or diffusive regime of conduction \cite{naidyuklibro,daghero10} is certainly likely to occur. A generalization of the BTK model for Andreev reflection in a diffusive contact has been developed by I.I. Mazin \emph{et al.} \cite{mazin01}. In a single-band superconductor, and even in the 1D case, this model predicts a sort of ``Z-enhancing effect'' with respect to the corresponding ballistic BTK model, which is somewhat similar to that due to the shape of the Fermi surface (e.g. for $c$ axis current injection). For example, a conductance curve that in the diffusive model is obtained with $Z=0$ has approximately the same shape of a curve that, in the ballistic BTK model, is obtained with $Z\simeq0.5$ \cite{mazin01}. On the basis of this simple result, it seems fairly unlikely that this model can remove or decrease the zero-bias discrepancy between theoretical and experimental spectra. Anyway, the combined use of a multi-gap diffusive model and of a realistic Fermi surface is  particularly interesting and will be the subject of future work.}}

The zero-bias anomaly seems therefore related to an additional enhancement of the PCARS conductance at low bias not included in the model. In order to better clarify this point, in Fig. \ref{fig:3DBTK}d we show the difference between the experimental curves and the 3D fit. Along the $ab$ plane this difference is negligible while in the $c$-axis direction a zero-bias conductance peak (ZBCP) is present that survives even at temperatures close to $T_c$. As shown in Fig. \ref{fig:3DBTK}d, at the increase of temperature this peak broadens with a strong reduction of its amplitude but an evident increase of its full width at half maximum (from 3.58 meV at 4.2 K to 3.86 meV at 19.3 K).

What is the possible origin of this ZBCP?  As already mentioned, the possible explanation in terms of Andreev bound states due to an unconventional nodal structure of (one of) the gaps seems to be ruled out by our specific calculations within the 3D BTK model. Another usual explanation of zero-bias peaks is the non-ballistic nature of the contact, but in our case this is ruled out by the clear presence of spectroscopic gap features at higher bias \cite{mondal12}. In our opinion, in this specific case the observed excess conductance might be ascribed to magnetic scattering in the contact region. This explanation might be reasonable because: i) in the phase diagram of Co-doped Ba-122, the region of coexistence of superconductivity and magnetism extends almost up to optimal doping \cite{lumsden10}; ii) in our samples, the probable inhomogeneity in the dopant content and the tendency to Co clustering may give rise to regions of different effective doping; iii) indeed, the local $T_c$ of the contact shown in Fig.\ref{fig:3DBTK} (22.5 K) is smaller than the bulk $T_c$ and corresponds to a local doping content around 6\%, just at the boundary with the AFM SDW region.
%

The observation of ZBCP explained in terms of Kondo resonance by diluted magnetic impurities at the N-I interface \cite{appelbaum67} of NIN' \emph{tunnel} junctions has been repeatedly reported in the past 50 years (for a very clear example see \cite{shen68}). More recently the Kondo effect has been observed by scanning tunneling spectroscopy in single atoms and molecules (for an example see \cite{komeda11} and references therein). In the present case, however, we are dealing with point contact that are \emph{not} in the pure tunneling regime. There is a large literature concerning the study of scattering by diluted magnetic impurities in point contacts (see \cite{naidyuklibro} and references therein); but if the contact is ballistic, the expected and observed behaviour at zero applied magnetic field (elastic spin-flip scattering) is a zero-bias \emph{depression} of the conductance and not a peak. ZBCPs have been instead observed in ballistic point contacts when the spin-flip scattering is inelastic, i.e. when an external magnetic field is present or the concentration of magnetic impurities is above a certain threshold so that an internal field originates from their interaction in the spin glass state \cite{duif89}.
A further element that should be borne in mind is the fact that the barrier height of our point-contact junctions is always greater than zero (usually between 0.2 and 0.5) thus allowing a certain degree of ``tunnel-like'' conduction which could make the results consistent with those observed in the pure tunnel regime.  Further experiments (and theoretical analyses) are thus necessary in order to show in a \emph{direct} way this ZBCP and study its behavior as a function of temperature, magnetic field and doping. Preliminary results of PCAR spectroscopy in slightly underdoped Co-doped Ba-122 thin films seem to clearly show the ZBCP thus confirming the picture discussed in the present section.

\section{Conclusions}
Since the discovery of Fe-based superconductors in 2008 our knowledge about these materials has impressively evolved. However, some very basic issues still remain unsolved, especially concerning the robustness of various properties over the different families; and new experiments aimed at addressing the open issues often instead unveil new and subtle problems.
In this paper we have tried to show that {\color{blue}{point-contact Andreev-reflection spectroscopy }} can be very helpful in the effort to answer some open questions, and we have presented a (necessarily partial) overview of recent advances in the field that, in our opinion, can lead to further developments. In particular, we have shown that PCARS possesses the capabilities to explore the whole phase diagram of FeSC, from the superconducting dome to the region of coexistence with magnetism (if present) and back to the undoped compounds. To do so, not only new measurements away from optimal doping are necessary, but also the development and the refinement of the theoretical tools used to interpret the experimental data {\color{blue}{in the superconducting state}}.

\section*{Acknowledgments}
Many thanks to G.A. Ummarino for his theoretical support in all these years and for kindly perusing the manuscript. A particular acknowledgment to V.A. Stepanov for for his fundamental contribution to our experimental research. Many thanks to all those who grew and characterized the samples: J. Karpinski, N. D. Zhigadlo, Z. Bukowski, J. Jiang, J. D. Weiss, E. E. Hellstrom, J. S. Kim, M. Putti, M. Tropeano and A. Palenzona. Special thanks to F. Bernardini, G. Profeta and S. Massidda, for enlightening discussions and invaluable theoretical contributions.
{\color{blue}{Finally, we acknowledge the financial support of the European Community through to the Collaborative EU-Japan Project "IRON SEA" (NMP3-SL-2011-283141)}}.

\bibliographystyle{apsrev}
\bibliography{bibliografia_def}


\end{document}